\documentclass[12pt]{article}
\usepackage[a4paper,left=30mm,right=20mm,top=20mm,bottom=20mm]{geometry}
\usepackage{epsfig}
\usepackage{xcolor}
\usepackage{graphics}
\usepackage{amsmath}
\usepackage{amssymb}

\newcommand{\bea}{\begin{eqnarray}}
\newcommand{\eea}{\end{eqnarray}}
\newcommand{\be}{\begin{equation}}
\newcommand{\ee}{\end{equation}}

\title{Shadowing and antishadowing in the rescaling model}

\author{A.V.~Kotikov$^{1,2}$, A.V.~Lipatov$^{2,3}$, P.M.~Zhang$^{1}$}

\begin{document}

\maketitle

\begin{center}
{\it $^{1}$School of Physics and Astronomy, Sun Yat-sen University, 519082 Zhuhai, China}\\
{\it $^{2}$Joint Institute for Nuclear Research, 141980 Dubna, Moscow region, Russia}\\
{\it $^{3}$Skobeltsyn Institute of Nuclear Physics, Lomonosov Moscow State University, 119991 Moscow, Russia}

\end{center}

\vspace{0.5cm}

\begin{center}

{\bf Abstract }
       
\end{center}

\indent
Nuclear deep inelastic structure functions $F_2^A(x,Q^2)$ as well as 
parton distribution functions in a nuclei are investigated in the framework of 
rescaling model. 
Our analysis is based on analytical expressions for quark and gluon densities in a proton
derived at the leading order of QCD coupling.
By fitting the rescaling parameters from
the experimental data on $F_2^A(x,Q^2)/F_2^D(x,Q^2)$ ratio 
for several nuclear targets, we derive 
predictions for corresponding nuclear parton distributions and, thus,
for shadowing and antishadowing effects.

\vspace{1.0cm}


\vspace{1.0cm}

\newpage 

\section{Introduction} \indent

The study of deep inelastic scattering (DIS) of leptons on nuclei
shows the appearance of a significant effect of nucleon interaction
in the nucleus, which eliminates the naive idea of the nucleus as
a system of quasi-free nucleons (see, for example, review~\cite{Arneodo:1992wf,Rith:2014tma}).
This effect was first discovered~\cite{Aubert:1983xm} by the
European Muon Collaboration (EMC)
in the domain of valence quark dominance, therefore it was called the EMC effect.
Influence of nuclear effects on parton distribution functions (PDFs) 
meets a lot of interest from both theoretical and experimental points of view.
In particular, detailed knowledge of PDFs in a nuclei (nPDFs) is necessary for any 
theoretical description of $pA$ and $AA$ 
processes studied at modern (LHC, RHIC) and future colliders (FCC-he, EiC, EicC, NICA).

Usually the nuclear modification factor, defined 
as a ratio of per-nucleon structure functions in nuclei $A$ and deuteron, $R = F_2^A(x,Q^2)/F_2^D(x,Q^2)$,
or rather ratio of corresponding parton densities,
is introduced and its behavior in the 
shadowing ($x \leq 0.1$), anti-shadowing ($0.1 \leq x \leq 0.3$), 
valence quarks and Fermi motion dominance regions ($0.3 \leq x \leq 0.7$ and $x \geq 0.7$, respectively) is investigated.
Unfortunately, up to now there is no commonly accepted
framework to describe this nuclear modification of PDFs
in a whole kinematical range.
Two main approaches are used by different groups at present.
In the first, which is currently seems to be more popular, nPDFs are extracted from a global
fit to nuclear data using some empirical
parametrization of corresponding intitial parton densities. 
Then, numerical solution
of Dokshitzer-Gribov-Lipatov-Altarelli-Parisi (DGLAP) equations is applied 
to describe their QCD evolution (see review~\cite{Paakkinen:2022qxn} and references therein).
The second strategy is based on special nPDF models (see,
for example,~\cite{Kulagin:2004ie,Jaffe:1983zw,Close:1983tn}
and review~\cite {Kulagin:2016fzf}).

The aim of this Letter is to study the nuclear modification of DIS structure 
function (SF) $F_2(x,Q^2)$ in the framework of the
rescaling model~\cite{Jaffe:1983zw}.
The rescaling model is
based on the assumption~\cite{Close:1983tn} that the effective size
of gluon and quark confinement in nucleus is greater than in a free nucleon. 
Within the framework of perturbative QCD 
this  confinement rescaling predicts~\cite{Jaffe:1983zw,Close:1983tn} that ordinary PDFs and nPDFs can be connected by simply
shifting the values of the kinematic variable $Q^2$  (see also review~\cite{Jaffe:2212}).
Thus, the rescaling model demonstrates the features
inherent in both approaches: there is the
relationship between PDFs and nPDFs that arises as a result of
shifting in the scale $Q^2$ and, at the same time, both PDFs and nPDFs
obey the DGLAP equations. In a sence, the rescaling model corresponds
to the first strategy with empirical nPDFs
obtained from the corresponding PDFs by $x$-independent shift.
Initially, it was proposed for the domain of valence quarks dominance
and expanded recently to a low $x$
range~\cite{Kotikov:2017mhk,Kotikov:2018ass}.

Our analysis is mainly based on the leading order (LO) analytical
expressions~\cite{Abdulov:2022itv}
for proton PDFs valid at both low and intermediate $x$ values.
At low $x$, these formulas were obtained in the so-called double asymptotic scaling (DAS) 
approximation \cite{Rujula,Q2evo}, which is related to the asymptotic
behaviour of DGLAP evolution \cite{DGLAP}.
Performing fit on the NMC data~\cite{Arneodo:1995cs}, we derive nPDFs
at low and moderate $x$ values ($x \leq 0.7$) for several nuclei targets
and then investigate shadowing and anti-shadowing effects\footnote{The investigations of 
shadowing and antishadowing effects (see~\cite{Stodolsky:1966am,Nikolaev:1975vy})
have been started before the EMC experimental data~\cite{EuropeanMuon:1988tpw} were appeared (see also~\cite{Nikolaev:1981dh} for overview).}
for DIS SF
$F_2(x, Q^2)$ 
itself and for corresponding quark and gluon distribution functions.
Main advantage of our approach is related with analytical 
solution of the DGLAP equations.

\section{Proton SF $F_2^{p}(x,Q^2)$, quark and gluon densities} \indent

It is well known that the proton SF $F_2^{p}(x,Q^2)$
at the leading order of QCD coupling 
can be presented in the simple form
\begin{equation}
F_2^{p}(x,Q^2) = \sum_{i=1} e_i^2  \left[f_{q_i}(x,Q^2) +
f_{\bar{q}_i}(x,Q^2) \right],
\label{S1.14}
\end{equation}
\noindent
where $f_{q_i}(x,Q^2)$ and $f_{\bar q_i}(x,Q^2)$ are the quark and antiquark 
densities in a proton, respectively.
In the four-flavor scheme (4FS), where $b$ and $t$ quarks are separated out
(which will always be used below), 
we
have \cite{Buras:1979yt}
\begin{equation}
F_2^{ep}(x,Q^2) = \frac{5}{18} \, f_{SI}(x,Q^2) + \frac{1}{6} f_{NS}(x,Q^2),
\label{S1.15}
\end{equation}
where the singlet part $f_{SI}(x,Q^2)$ contains the valence and sea quark parts
$f_{V}(x,Q^2)$ and $f_{S}(x,Q^2)$:
\be
f_{V}=
f_u^V + f_d^V, \quad
f_{S} = \sum_{i=1}^{4} \left(f_{q_i}^S + f_{\bar{q}_i}^S \right), \quad
f_{SI}=
\sum_{i=1}^{4} \left(f_{q_i} + f_{\bar{q}_i} \right) = f_{V} + f_{S}.
\label{S1.12}
\ee
\noindent
The nonsinget part $f_{NS}(x,Q^2)$ contains difference between the up and
down quarks:
\be
f_{NS}
= \sum_{q=u,\,c} \left[f_{q}(x,Q^2) +
f_{\bar{q}}(x,Q^2) \right]
- \sum_{q=d,\,s} \left[f_{q}(x,Q^2) +
f_{\bar{q}}(x,Q^2) \right].
\label{S1.16}
\ee
\noindent
Using the $Q^2$-evolutions for large and small values of $x$ (see~\cite{Gross,LoYn,Rujula,Q2evo,Kotikov2007}),
the analytical expressions~\cite{Abdulov:2022itv} for PDFs were 
constructed\footnote{In this sense, the study~\cite{Abdulov:2022itv} is a continuation of previous investigations~\cite{Illarionov:2010gy} carried out for valence quarks.}.

\subsection{Nonsinglet and valence parts} \indent

The nonsinglet and valence quark part $f_i(x,Q^2)$, where $i = V$ or $NS$, can be represented in the following form
(see \cite{Abdulov:2022itv})
\begin{equation}
f_i(x,Q^2) = 
\left[A_i(s)x^{\lambda_i}(1 -x) + \frac{B_i(s)\, x}{\Gamma(1+\nu_i(s))} + D_i(s)x (1 -x) \right] (1-x)^{\nu_i(s)},
\label{S4.1_l}
\end{equation}
\noindent
where $\Gamma$ is the Riemann's $\Gamma$-function,
\bea
&&\nu_{i}(s)=\nu_{i}(0)+r s,~~B_{i}(s)=B_{i}(0) e^{-p s},~~p=r\bigl(\gamma_{\rm E}+\hat{c}\big),~~
r=\frac{16}{3\beta_0},~~\hat{c}=-\frac{3}{4}, \nonumber \\
&&A_{i}(s)=A_{i}(0) e^{-d(1-\lambda_i)
  s},~~d(1-\lambda_i)=\frac{\gamma_{NS}(1-\lambda_i)}{2\beta_0},~~s = \ln \left[{\alpha_s(Q_0^2)\over \alpha_s(Q^2)}\right].
\label{A_i}
\eea
and $\gamma_{\rm E} \simeq 0.5772$ is the Euler's constant,
$\beta_0=11-(2/3)N_f$ is the LO
QCD $\beta$-function,
$N_f = 4$ is a number of active (massless) quarks 
and $\gamma_{NS}(n)$ is the LO NS anomalous dimension. 
The factors $Q_0^2$, $A_i(0)$, $B_{i}(0)$ and $\nu_{i}(0)$ are free
parameters which can be determined from experimental
data\footnote{For example, from precision HERA data on proton structure function $F_2(x,Q^2)$.}.

We note that~(\ref{S4.1_l})
is constructed as a combination of the small-$x$
part proportional to $A_i(s)$, 
large-$x$ asymptotics proportional
to $B_i(s)$ and 
additional term $\sim D_i(s)$, which is
subasymptotics in both the regions.
Dependence on scale $Q^2$ in~(\ref{S4.1_l}) is given by~(\ref{A_i}).
The $Q^2$-dependence of the magnitude  $D_i(s)$
is determined by corresponding sum rules (see~\cite{Abdulov:2022itv}).
Everywhere below, we apply "frozen" treatment 
of the QCD coupling (see, for example, \cite{Kotikov:2004uf} and
references therein), where $\alpha_s(Q^2) \to \alpha_s(Q^2 + M_\rho^2)$
with $M_\rho \sim 1$~GeV, that immediately leads to $s \geq 0$.



\subsection{Sea and gluon parts} \indent

It was shown~\cite{Abdulov:2022itv} that
the sea and gluon parts can be represented as combinations\footnote{The full set of formulas is listed~\cite{Abdulov:2022itv}. 
Here we omit the large $x$ contribution for the "$+$" component, which is negligible in comparison
with the "$-$" and valence component. Moreover, from fits of
experimental data we found~\cite{Abdulov:2022itv} that  the large $x$
contribution for the "$-$" component is also negligible in comparison
with the corresponding contribution of the valence quarks.} of the "$\pm$" parts:
\begin{gather}
   f_i(x,Q^2) = \sum_{\pm} f_i^{\pm}(x,Q^2),
\end{gather}
\noindent
where $i = S$ or $g$ and
\begin{gather}
  f_{S}^-(x,Q^2) = \left[A_{q}e^{- d_{-} s} (1-x)^{m_{q,-}} + D_{-}(s)x (1 -x) \right] (1-x)^{\nu_{-}(s)},
  \label{f1}
\end{gather} 
\begin{gather}
  f_{S}^+(x,Q^2)= {N_f\over 9} \left(A_g + {4\over 9}A_q\right) \rho I_1(\sigma)e^{-\overline d_{+} s} (1-x)^{m_{q,+}+\nu_{+}(s)+1}\,,
\end{gather}
\begin{gather}
  f_{g}^-(x,Q^2) = - {4\over 9} A_q e^{- d_{-} s} (1-x )^{m_{g,-}+\nu_{-}(s)+1} + \nonumber \\
    + \frac{K_{-}}{\Gamma(2+\nu_{V}(s))}  \, \frac{B_V(s) x (1-x)^{\nu_{V}(s)+1}}{\left[\ln(1/(1-x))+\hat{c} + \Psi(\nu_{V}(s)+2)\right]}, 
\end{gather} 
\begin{gather}
  f_{g}^+(x,Q^2) =  \left(A_g + {4\over 9}A_q\right) I_0(\sigma)e^{-\overline d_{+} s} (1-x)^{m_{g,+}+\nu_{+}(s)}.
  \label{f2}
\end{gather} 
\noindent
Here $\Psi$ is the Riemann's $\Psi$-function and
\bea
\nu_{\pm}(s)=\nu_{\pm}(0)+r_{\pm}s,~~B_{\pm}(s)=B_{\pm}(0)
e^{-p_{\pm}s},~~p_{\pm}=r_{\pm}\bigl(\gamma_{\rm E}+\hat{c}_{\pm}\big), \nonumber \\
r_{+}=\frac{12}{\beta_0},~~r_{-}=\frac{16}{3\beta_0},~~\hat{c}_{+}=-\frac{\beta_0}{12},~~\hat{c}_{-}=-\frac{3}{4},~~K_{+}=\frac{3N_f}{10},~~K_{-}=\frac{2}{5}\,.
\label{rcpm}
\eea
\noindent
Here $I_{0}$ and $I_1$ 
are the modified Bessel functions
with
\be
\sigma = 2\sqrt{\left|\hat{d}_+\right| s
  \ln \left( \frac{1}{x} \right)},~ \rho=\frac{\sigma}{2\ln(1/x)},~
  \hat{d}_+ = - \frac{12}{\beta_0},~
\overline d_{+} = 1 + \frac{20N_f}{27\beta_0},~
d_{-} = \frac{16N_f}{27\beta_0}.
\label{intro:1a}
\ee
The factors 
$A_g$, $A_q$, $B_{\pm}(0)$ and $\nu_{\pm}(0)$ are 
taken to describe the HERA data on proton SF $F_2(x,Q^2)$.
The $Q^2$-dependence of the subasymptotic terms $\sim D_{\pm}(s)$ is taken from the momentum conservation law.
Following~\cite{Abdulov:2022itv}, we set $m_{q,-}=m_{g,+}=2$ and $m_{q,+}=m_{g,-}=1$. In this case, the small $x$ asymptotics are suppressed at large $x$ compared to
the subasymptotic terms $\sim D_{\pm}(s)$.
Moreover, the small $x$ asymptotics contain the same powers of $(1-x)$ factor for quarks and gluons.\\

\section{Rescaling model at large and low $x$} \indent

According to
the rescaling model~\cite{Jaffe:1983zw}, the DIS
structure function
$F_2^p(x,Q^2)$ and, consequently, the valence and nonsinget parts of quark densities 
in the case of
a nucleus $A$ are modified 
at intermediate and large $x$ 
as follows\footnote{We use the same shift $Q^2_{A,V}$ for both the valence and nonsinget parts.}
\begin{equation}
  f_{i}^A(x,Q^2) =
  f_{i}(x,Q^2_{A,V}),
  \label{va.1}
\end{equation}
where $i = V$ or $NS$ and new scale $Q^2_{A,V}$ is related to $Q^2$ by \cite{Kotikov:2017mhk}
\begin{equation}
s^A_V \equiv \ln \left(\frac{\ln\left(Q^2_{A,V}/\Lambda^2_{\rm QCD}\right)}{\ln\left(Q^2_{0}/\Lambda^2_{\rm QCD}\right)}\right)
= s +\ln\Bigl(1+\delta^A_V\Bigr) \approx s +\delta^A_V,~~~
\label{sA}
\end{equation}
so the kernel modification of main variable $s$ depends on the
$Q^2$-independent parameter $\delta^A_V$ having small values (see~\cite{Kotikov:2017mhk}).
Then, 
the small $x$ PDF asymptotics 
have been applied~\cite{Kotikov:2017mhk} to the small $x$ region of the EMC effect using the simple fact
that the rise of sea quark and gluons densities increases with increasing $Q^2$.
Thus, in the case of nuclei, the PDF evolution scale is less than $Q^2$ and this can directly reproduce the shadowing effect observed in global fits.
Since there are two components for each parton density, "$+$" and "$-$" ones, 
we have two free parameters $Q^2_{A,\pm}$ that can be determined from the analysis of
experimental data for the EMC effect at low $x$ values.
Usually, it is convenient to study the following ratio
\begin{equation}
R^{AD}_{i}(x,Q^2) = \frac{f^A_i(x,Q^2)}{f^D_i(x,Q^2)},
\label{AD}
\end{equation}
\noindent
where $i = S$, $SI$ or $g$. 
Taking advantage of the fact that the nuclear effect in the deuteron is very 
small,
$f^D_i(x,Q^2) \approx f_i(x,Q^2)$, we can assume\footnote{Study of nuclear effects in the 
deuteron can be found~\cite{AKP,Wang:2016mzo,AbdulKhalek:2022fyi}. See also discussions~\cite{Kulagin:2016fzf}.} that
\be
F^A_2(x,Q^2) =
\frac{5}{18}\,
  f^{A}_{SI}(x,Q^2) + \frac{1}{6}\,
  f^{A}_{NS}(x,Q^2),~~
  f^{A}_{SI}(x,Q^2)=
f^{A}_S(x,Q^2)+f^{A}_V(x,Q^2)\,,
\ee
where $f^{A}_V(x,Q^2)$ and $f^{A}_{NS}(x,Q^2)$ are
given by~(\ref{va.1}).
The gluon and sea quark densities are calculated as combinations of "$\pm$" components:
\be
f_a^{A,+}(x,Q^2)
~=~
f_a^{A,+}(x,Q^2) + f_a^{A,-}(x,Q^2),~~
f^{A,\pm}_a(x,Q^2) =
f^{\pm}_a(x,Q^2_{AD,\pm}) \, .
\label{AD1}
\ee
\noindent
with $a = g$ or $S$.
The expressions for $f^{\pm}_a(x,Q^2)$ are given by~(\ref{f1}) --- (\ref{f2}) and 
corresponding values of $s^{AD}_{\pm} $ turned out to be
\be
s^{AD}_{\pm} \equiv \ln \left(\frac{\ln\left(Q^2_{AD,\pm}/\Lambda^2_{\rm QCD}\right)}{\ln\left(Q^2_{0}/\Lambda^2_{\rm QCD}\right)}\right)
  = s +\ln\Bigl(1+\delta^{AD}_{\pm}\Bigr)\,,
\label{AD2}
\ee
where $\delta^{AD}_{\pm}$ can be obtained by fitting
the NMC experimental data
on structure function ratios $F_2^A(x,Q^2)/F_2^D(x,Q^2)$~\cite{Arneodo:1995cs}.

. 

\section{Numerical results} \indent

\begin{figure}
\includegraphics[width=5.2cm]{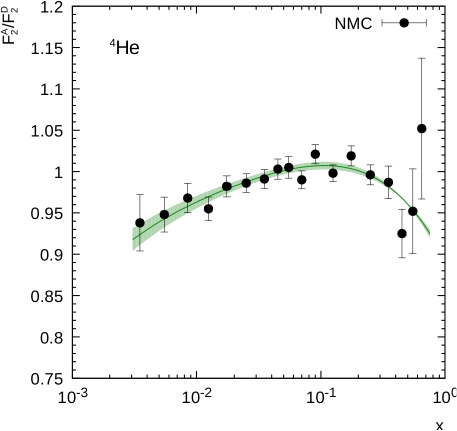}
\includegraphics[width=5.2cm]{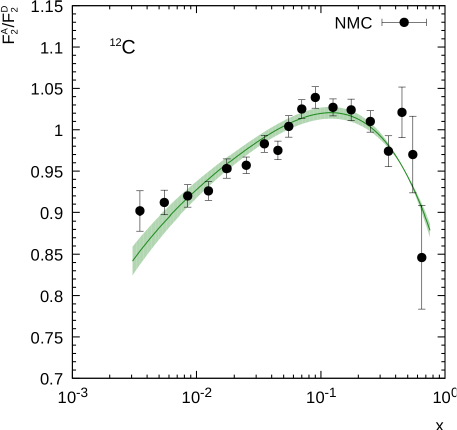}
\includegraphics[width=5.2cm]{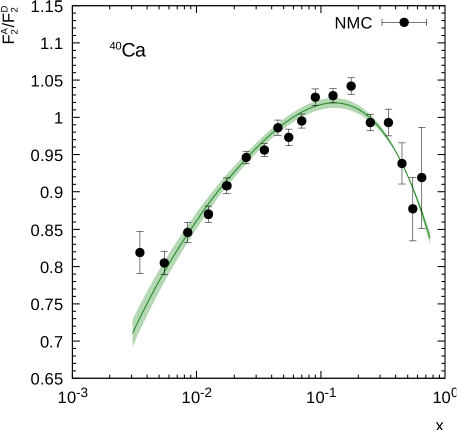}
\caption{The fit results of structure function ratios
$F_2^A(x,Q^2)/F_2^D(x,Q^2)$ for several nuclear targets 
compared to the NMC experimental data~\cite{Arneodo:1995cs}.}
\end{figure}

Performing fits on the NMC data~\cite{Arneodo:1995cs} 
at low and moderate $x$ values, $x \leq 0.7$, we obtained the values of
$\delta^{AD}_{\pm}$ for several nuclei targets.
Our results are shown in Fig.~1 and collected in Table~1,
where we additionaly show the $\delta^{AD}_{V}$
values taken from~\cite{Kotikov:2017mhk}.
The goodness of our fit, $\chi^2/n.d.f.$, is presented also.
One can see that newly fitted $\delta^{AD}_{+}$ are about two times less
than obtained earlier~\cite{Kotikov:2017mhk}.
Moreover, we find that derived values of $\delta^{AD}_{-}$ differ in sign from 
the previous results~\cite{Kotikov:2017mhk}.
The main sourse of this difference is 
that the small-$x$ PDF asymptotics 
have been used in the analysis~\cite{Kotikov:2017mhk} and, therefore,
the NMC data were considered at low $x$ only.
In contrast, here we extended the consideration
into the region of moderate $x$ and take into account all NMC data points.
Neverveless, strong difference in fitted $\delta^{AD}_{\pm}$
leads to not so different results 
for nuclear modification factor $R$
at low
$x$ values. It is because $Q^2$-changes in the "$+$" and "$-$" components
occur in opposite directions: in fact, the "$+$" ("$-$") component increases
(decreases) with by increasing $Q^2$.
So, for the values of $\delta^{AD}_{\pm}$ derived in previous analysis~\cite{Kotikov:2017mhk}, 
the contribution of the "$-$" components
is the opposite of the contribution of the "$+$" components.
For the values of $\delta^{AD}_{\pm}$ collected in Table~1, the contributions
of the two components reinforce each other.

\begin{table} 
\begin{center}
\begin{tabular}{|c|c|c|c|c|c}
\hline
$A$ &  $^4$He & $^{12}$C &  $^{40}$Ca    \\
\hline
$\delta^{AD}_{+}$   & $-0.021 \pm 0.003$ & $ -0.040 \pm 0.003$  & $-0.067 \pm 0.003$   \\
$\delta^{AD}_{-}$   & $0.13 \pm 0.02$ &  $0.26 \pm 0.04$ & $0.35 \pm 0.04$   \\
$\delta^{AD}_V$   & $0.05$   & 0.07 & 0.10  \\
\hline
$\chi^{2}/n.d.f.$   & $0.74$                &  $1.75$            & $1.94$   \\
\hline 
\end{tabular}
\caption{The $\delta^{AD}_{\pm}$ parameters extracted from the NMC data~\cite{Arneodo:1995cs} on structure 
function ratios $F_2^A(x,Q^2)/F_2^D(x,Q^2)$. The $\delta^{AD}_{V}$ values are
taken from~\cite{Kotikov:2017mhk}.} 
\end{center}
\end{table}

Since at low $x$ the "$+$" component increases strongly with $Q^2$
growth, it is mainly responsable for the shadowing effect. For antishadowing, the
contribution of the "$-$" component is more important. Strictly
speaking, from momentum conservation in the case of a nuclear target\footnote{$\delta^{AD}_{V} \approx \delta^{A}_{V}-\delta^{D}_{V}$ with
$\delta^{D}_{V} \approx 0.1$, see~\cite{Kotikov:2017mhk}.} one obtain
$\delta^{AD}_{-}=\delta^{AD}_{V}$. However,
we cannot use momentum conservation here since we do not
consider the large $x$ range, $x \geq 0.7$, where the Fermi motion should
be taken into account.

Next, using the analytical expressions (\ref{S4.1_l}) --- (\ref{intro:1a}) for nucleon target,
(\ref{va.1}) --- (\ref{AD2}) for nuclear targets and fitted
values of $\delta^{AD}_{\pm}$, 
we can give
predictions for nuclear modification factors $R^{AD}_{i}(x,Q^2)$ defined by~(\ref{AD}).
Our results are shown in Fig.~2 for several nuclear
targets, namely, $^4$He, $^{12}$C and $^{40}$Ca. Since, as is well known,
$R^{AD}_{i}(x,Q^2)$ are practically independent on $Q^2$, except may be low
$x$ range, where, however, $Q^2$-dependence is not so large, too. To save space,
we show the results for $Q^2 = 10$~GeV$^2$ only.
We find that the shadowing effect for gluons is less than for quarks,
that is consistent with the results of other studies~\cite{Wang:2016mzo,AbdulKhalek:2022fyi,Eskola:2021nhw,Hirai:2007sx}.
Shadowing for antiquarks and quarks (sea and singlet quark densities, respectively) 
is very similar for $x\sim 10^{-3}$, but a bit stronger for
antiquarks at $x\sim 10^{-3}$, that is in full agreement with predictions~\cite{Wang:2016mzo}.
Gluon antishadowing is absent, which is also consistent with~\cite{Wang:2016mzo}, 
where that antishadowing effect is very small. However, it is in disagreement with predictions
of other groups~\cite{AbdulKhalek:2022fyi,Eskola:2021nhw}, where antishadowing has a great effect on gluons.
The antishadowing for antiquarks is greater than for  valence quarks, that 
contradicts the results of the study~\cite{Wang:2016mzo}. Other groups present results for 
antiquark antishadowing with large uncertainties, and therefore, it is 
difficult to draw any specific conclusion at present.

\begin{figure}
\includegraphics[width=5.2cm]{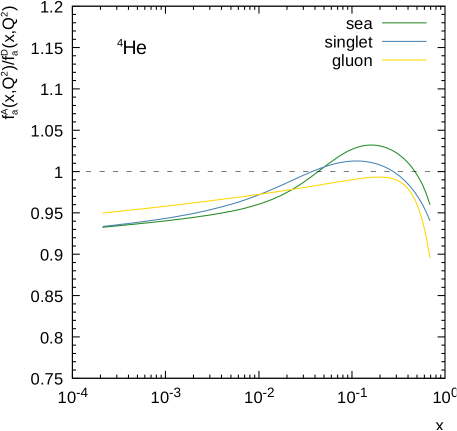}
\includegraphics[width=5.2cm]{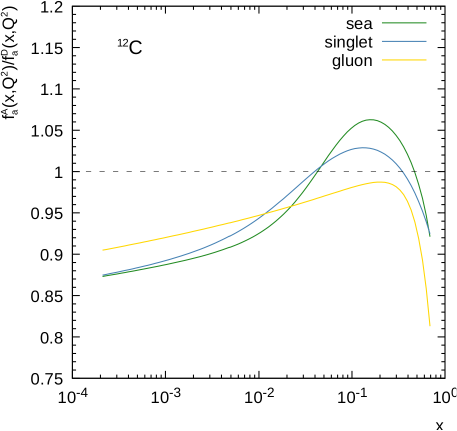}
\includegraphics[width=5.2cm]{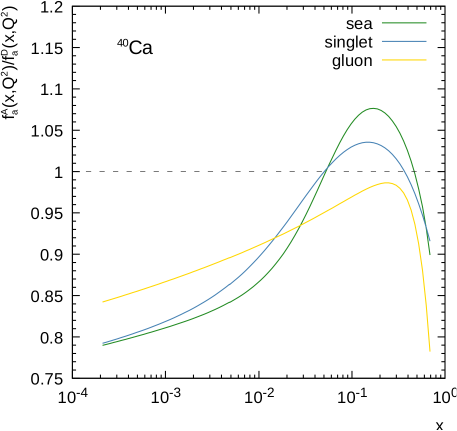}
\caption{The predicted nuclear modification factors for parton distributions in several nuclear targets. Fixed value $Q^2 = 10$~GeV$^2$ 
is applied.}
\end{figure}

\section{ Conclusion} \indent

In the framework of rescaling model, we fitted the NMC experimental data
for the ratios of the DIS structure functions $F_2(x,Q^2)$ in nuclear targets and deuteron
at low and intermediate $x$ values, $x \leq 0.7$.
Our analysis is based on the analytical expressions for proton PDFs
derived previously in \cite{Abdulov:2022itv}.
Using the obtained resuls for rescaling values, we derive 
predictions for nPDFs for several nuclear targets and, thus,
for shadowing and antishadowing effects.
We find that shadowing effect for gluons is less than for quarks,
which is consistent with many other studies. There is no antishadowing
for gluons, and it is better pronounced for antiquarks than for quarks. This is
a rather interesting result, since different groups give very
different results on the antishadowing effect with large uncertainties. 

As the next steps, we plan to include the Fermi motion in our consideration
and derive results for nuclear modifications of parton densities
over the entire $x$ range. Moreover, we plan to study nuclear modifications
of Transverse Momentum Dependent parton distribution functions
\cite{Kotikov:2019kci,Abdulov:2023kfp}, 
which are now become very popular (see \cite{Abdulov:2021ivr})
in the phenomenological analyses.

\section*{Acknowledgements} \indent

A.V.K. and A.V.L. would like to thank School of Physics and Astronomy,
Sun Yat-sen University (Zhuhai, China) for warm hospitality.
P.M.Z. was partially supported by the National Natural Science
Foundation of China (Grant No. 12375084).



\end{document}